# Entangled photons on-demand from single quantum dots

**Robert J. Young[1], R. Mark Stevenson[1], Paola Atkinson[2], Ken Cooper[2], David A. Ritchie[2], Andrew J. Shields[1]**

[1]Toshiba Research Europe Limited, 260 Cambridge Science Park, Cambridge CB4 0WE, UK.

[2]Cavendish Laboratory, University of Cambridge, Madingley Road, Cambridge CB3 0HE, UK.

E-mail: Robert.young@crl.toshiba.co.uk

**Abstract.** We demonstrate the on-demand emission of polarisation-entangled photon pairs from the biexciton cascade of a single InAs quantum dot embedded in a GaAs/AlAs planar microcavity. Improvements in the sample design blue shifts the wetting layer to reduce the contribution of background light in the measurements. Results presented show that >70% of the detected photon pairs are entangled. The high fidelity of the $(|H_{XX}H_X\rangle+|V_{XX}V_X\rangle)/\sqrt{2}$ state that we determine is sufficient to satisfy numerous tests for entanglement. The improved quality of entanglement represents a significant step towards the realisation of a practical quantum dot source compatible with applications in quantum information.

A source of entangled photon pairs is a vital commodity for quantum information applications based on quantum optics[1], such as entanglement based protocols for quantum key distribution[2], long distance quantum communication using quantum repeaters[3], and to realise an optical quantum computer[4]. For all these applications, the number of photon pairs generated per cycle is of critical importance, since emission of multiple photon pairs introduces errors due to the possibility that two individual photons are not entangled. The most widely used technique to generate entangled photon pairs is currently parametric down conversion[5,6], which produces a probabilistic numbers of photons pairs per excitation cycle. In contrast, the biexciton decay in a single quantum dot was proposed to provide a source of 'triggered' entangled photon pairs, so called because it can produce no more than two photons per excitation cycle[7]. Such a device could be a favourable alternative to parametric down converters for future applications in quantum optics, with the added benefit that it might be realised in a simple structure similar to an LED[8]. Until recently, the realisation of such a device has been prevented due to polarisation splitting of the dot emission lines. The 'which-path' information that this provides destroys any entanglement, resulting in only classically correlated entangled photon pairs[9-11]. However, by controlling the polarisation splitting using growth[12] or magnetic fields[13], we recently demonstrated for the first time that triggered entangled photon pairs are emitted by dots with zero polarisation splitting[14]. Many challenges still remain however, in order to realise a practical quantum dot source of entangled photons. Improvements must be made to the efficiency of the device, the frequency of operation, and most importantly the degree of entanglement. In this paper we present results from quantum dots incorporated into an improved sample design which suppresses the amount of background light detected by our experiments. This approach allows us to measure more than three times as much entanglement than we previously reported[14], with a fidelity for the entangled $(|H_{XX}H_X\rangle+|V_{XX}V_X\rangle)/\sqrt{2}$ state we measure of 0.702 ± 0.022.

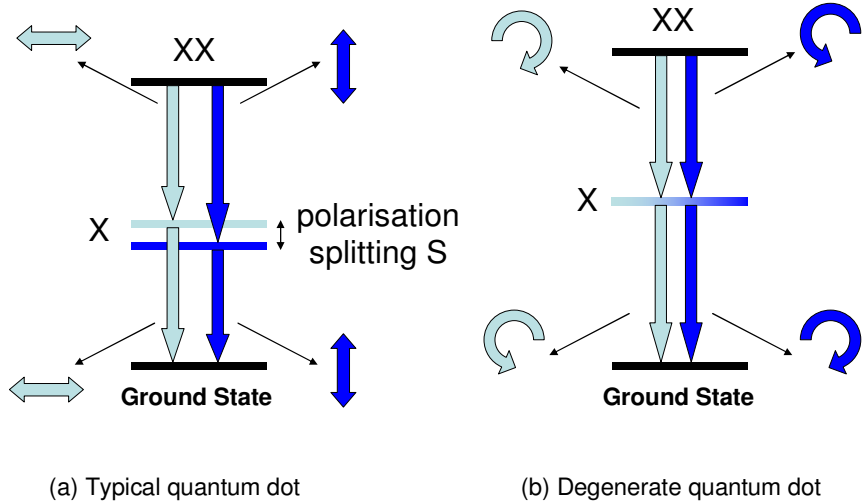

**Figure 1:** Schematic showing the radiative decay of the biexciton state (XX) in (a) a typical quantum dot, and (b) a quantum dot with zero splitting $S$ of the intermediate exciton level. For typical quantum dots, the radiative decay of XX generates a pair vertically or horizontally co-linearly polarised photons. For dots with zero polarisation splitting, the photon pairs emitted are super-positions of cross-circularly polarised pairs, and are entangled.

The emission of a pair of photons from a quantum dot is shown schematically in figure 1, and begins with excitation of the biexciton state XX. The biexciton state consists of paired electrons and heavy holes with opposing spin, and is spin neutral, along with the ground state. The spin dependent properties of the emission such as polarisation, and polarisation dependent energy splitting, are therefore determined by the intermediate exciton state X. In ordinary quantum dots, such as depicted by figure 1(a), structural properties of the dot such as elongation and strain cause in-plane asymmetry of the exciton wavefunction, which results in the hybridisation and energy splitting of the optically active exciton spin states via the exchange interaction[9-11,15,16]. The splitting $S$ of the exciton level allows the polarisation of each photon to be determined by energy measurements, which represents a type of 'which-path' information that destroys entanglement. The removal of the intermediate exciton level splitting is therefore crucial in order to realise triggered entangled photon pair emission from a quantum dot, as shown in figure 1(b).

We have previously presented the first demonstration that quantum dots can be engineered with fine structure exciton splitting less than the ~1.5μeV homogeneous linewidth of the exciton transition[12]. This was achieved by optimising the growth conditions of the quantum dots to minimise the splitting. It is also possible to reduce the splitting of some quantum dots by the application of modest in-plane magnetic fields[13]. Common to both approaches is the requirement that the quantum dots have rather high emission energy of at least 1.4 eV. This is because a dependence exists between the fine structure splitting $S$ and the exciton emission energy, attributed to the changing in-plane confinement of the exciton. The only dots suitable for entanglement are those emitting at ~1.4eV, which have zero splitting, and those emitting >1.4eV, which have inverted polarisation splitting that can be cancelled by in-plane magnetic fields. A limitation of these approaches is that the wetting layer emission is at ~1.42eV, with linewidth of ~10meV, which results in background light emission at 1.4 eV with comparable intensity to that of a dot. As a result, in ref [14], 49% of the coincident counts were due to background light emission, which strongly dilutes the light emitted by the quantum dot, ultimately limiting the observability of entanglement.

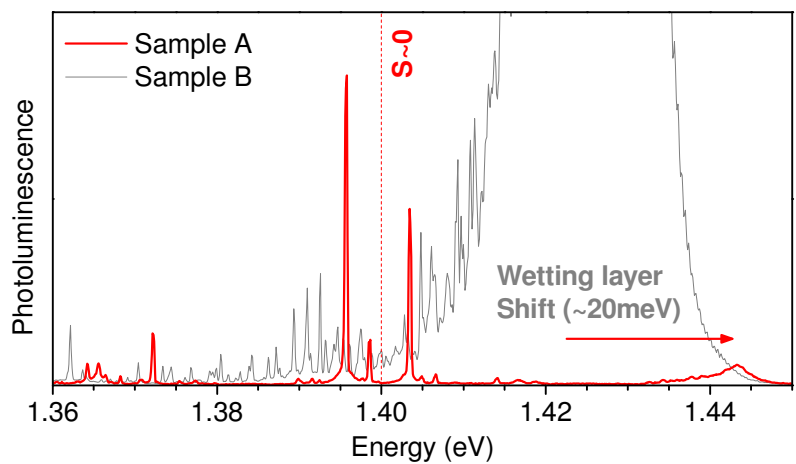

**Figure 2:** Micro-photoluminescence from an unprocessed area of the InAs quantum dot sample A used in these experiments. The sample incorporates a planar microcavity designed to increase the collection efficiency at 1.4 eV, the energy at which the dot polarisation splitting S is close to zero. To demonstrate the effect of the microcavity and the ~20°C increase in growth temperature, PL from a second sample containing only a typical layer of dots is shown in grey.

To suppress the background light levels we modified the sample design to blue shift the wetting layer emission away from the quantum dot emission. This was achieved by increasing the growth temperature by nominally 20°C, to encourage intermixing of the InAs wetting layer with the surrounding GaAs. The sample was grown by molecular beam epitaxy on a GaAs substrate, and included a single self-assembled quantum dot layer, with the thickness of InAs increased in response to the higher temperature, optimised to achieve the desired quantum dot density of ~$1\mu m^{-2}$. AlAs/GaAs distributed Bragg reflectors were grown above (2 repeats) and below (14 repeats) the dot layer to form a planar microcavity, resonant with the optimum quantum dot energy of 1.4eV, which enhances the light collection efficiency from the top of the sample by an order of magnitude[17,18].

Photoluminescence (PL) was measured at ~10K, with excitation provided non-resonantly using a 635nm laser diode emitting 100ps pulses with an 80MHz repetition rate. A microscope objective lens focussed the laser onto the surface of the sample, and collected the emitted light. In Figure 2, we show PL measured from an un-processed area on this sample (A), and for a sample grown cooler, and without a cavity (B), for comparison. For both measurements, a number of sharp lines are observed, from multiple quantum dots, together with a broad feature corresponding to the 2D wetting layer. Emission from the wetting layer is blue-shifted by ~20 meV for sample A, as a result of GaAs intermixing with the InAs. The effect of the cavity is two-fold. First, there is a notable enhancement in the PL collected from the quantum dots in sample A due to resonance with the optical cavity. Second, the wetting layer emission for sample A is suppressed by the stop-band of the cavity. Most importantly, the background light levels are greatly suppressed relative to the dot emission for sample A. Note that without blue shifting the wetting layer, the background intensity is unchanged relative to the dot, even with a cavity, since background emission resonant with the dot is equally enhanced.

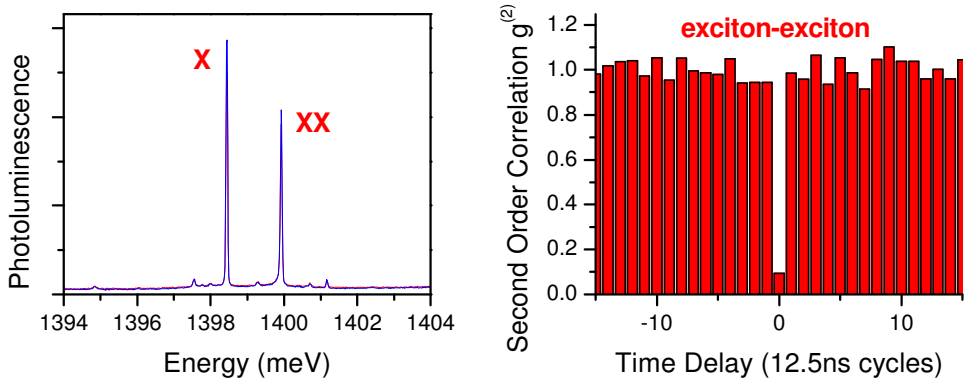

**Figure 3:** Role of background in measurements of single quantum dot emission. (a) shows the spectra of a single quantum dot, dominated by the neutral biexciton (XX) and exciton (X) emission lines. The intensity of the background emission is low. (b) Second order correlation between vertically and horizontally polarised X emission from a quantum dot. The strong suppression of coincidences at zero time delay indicates low background light levels.

To isolate single quantum dots, a metal shadow mask containing circular apertures around 2μm in diameter was defined on the surface of the sample. Emission lines corresponding to the neutral biexciton and exciton decay from single dots (as shown in figure 3a) are in-homogeneously broadened by fluctuating local charge distributions, and have measured line widths of ~50μeV. However, by fitting the XX and X vertically and horizontally polarised emission lines, it is possible to determine the linear polarisation splitting to within ~0.5μeV, as documented elsewhere[12]. Even with the blue shifted wetting layer, the relationship between splitting and emission energy for sample A is remarkably similar to that of sample B. This is unexpected, since the higher energy wetting layer must affect the in-plane confinement of the quantum dots, which in turn would modify the energy for which the splitting is zero. We speculate that the tall height of our quantum dots (~6nm) with respect to the wetting layer renders the precise vertical composition profile of the wetting layer relatively unimportant. Thus by characterising those quantum dots on sample A that emit around 1.4 eV, it was possible to select a quantum dot with approximately zero splitting, and additionally, low background contribution.

To analyse the properties of photon pairs emitted by a selected quantum dot, we measure the polarisation and time dependent correlations between the XX and X photons. Emission corresponding to the first and second photon emitted by the radiative decay of the biexciton was isolated by two spectrometers, tuned to the XX and X emission energies respectively. The insertion of appropriately oriented quarter-wave or half-wave plates preceding each of the two spectrometers allows any polarisation measurement basis to be selected. The spectrally filtered XX, and X emission passed through a linear polariser, and polarising beam splitter respectively, and was detected by three thermo-electrically cooled silicon avalanche photo-diodes (APDs). The time intervals between detection events on different APD's were measured, to determine the second order correlation functions. Finally, the number of counts was integrated over each quantum dot decay cycle.

To assess the role of background light, we first consider the correlation between the two orthogonally polarised components of the X emission. Since the radiative decay of the biexciton state can produce only one exciton photon, the probability of detecting two exciton photons simultaneously should be zero. Figure 3(b) shows an example of such a correlation, between horizontally and vertically polarised X emission. Each bar is proportional to the number of photon pairs detected, separated in time by the number of excitation cycles as shown on the bottom axis. The dip at zero delay is just 9.2 ± 4% of the average of the other peaks; corresponding to a suppression in the probability of emitting multiple exciton photons by an order of magnitude. The residual multiple exciton emission probability is a measure of the proportion of background light entering the detection system, and agrees well with the background levels measured directly from the photoluminescence spectra. We determine that background light is likely to contribute at least ~14% of the coincident counts in cross correlation measurements. This represents a more than three fold improvement over the 49% background coincidences observed previously.

For cross correlations between XX and X, the probability of detecting a pair of coincident photons, relative to the probability of detecting photons separated by a number of excitation cycles, is proportional to the inverse of the probability of generating an X photon per cycle. Thus, the relative probability of detecting coincident photons is dependent on the excitation rate, which fluctuates during the integration time of our experiments. However, the correlations of the XX detection channel with each of the orthogonally polarised X detection channels are measured simultaneously with the same excitation conditions, and thus can be compared directly. Additionally, we verify that the time averaged XX and X emission from this quantum dot is unpolarised within experimental error, so the number of coincident photon pairs can be normalised relative to the average number of photon pairs separated by at least one cycle, which compensates for the different detection efficiencies of the measurement system. We consequently define the degree of polarisation correlation $C$ according to equation 1, where $g_{XX,X}$ and $g_{XX,\bar{X}}$ are the simultaneously measured, normalised coincidences of the $XX$ photon with the co-polarised $X$ and orthogonally-polarised $\bar{X}$ photons respectively.

$$C = \frac{g_{XX,X} - g_{XX,\bar{X}}}{g_{XX,X} + g_{XX,\bar{X}}}$$

**Equation 1**

The degree of polarisation correlation $C$ varies between -1 and 1, where ±1 represents perfect polarisation correlation, and 0 represents no polarisation correlation. This measure will be used throughout the rest of this manuscript, as it represents the lowest error measurement for our system.

Figure 4(a) shows the degree of correlation between H polarised XX and H and V (rectilinearly) polarised X photons, as a function of the time delay between the photons, in cycles. For non-zero time delays, $C\sim0$, indicating that there is no polarisation memory from one cycle to the next. The noise on the data is caused by the statistical errors associated with the finite number of counts, from which the error $\Delta C$ can be estimated to be typically <0.05. At zero time delay, the large peak of 70% demonstrates a strong polarisation correlation between H polarised excitons and biexcitons in this dot.

In fact this high degree of polarisation correlation is present for any co-linearly polarised measurement bases for this dot. Figure 4(b) shows a similar correlation of diagonally (D) polarised biexcitons with diagonally polarised excitons of ~61%. Figure 4(d) plots the degree of correlation as the function of the angle of a single half wave plate, placed directly after the microscope objective. It is found that the degree of polarisation correlation $C$ is approximately independent of the half wave plate angle. This is an expected result for photon pairs being emitted in the entangled $(|H_{XX}H_X\rangle+|V_{XX}V_X\rangle)/\sqrt{2}$ state, since the linear polarisation measurement of the first photon defines the linear polarisation of the second

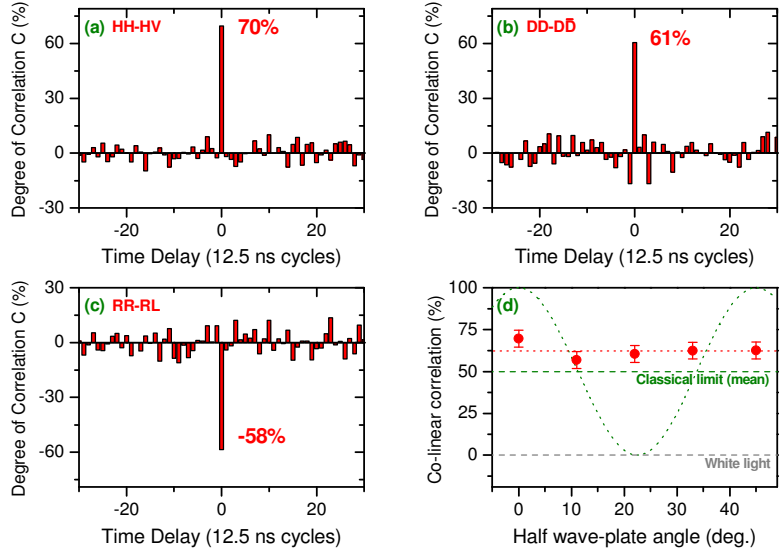

**Figure 4:** Biexciton – exciton polarisation correlations for a quantum dot with exciton level splitting $S\sim 0$. The degree of polarisation correlation $C$ is defined as the difference between the normalised polarisation correlated and anti-correlated coincidences, divided by their sum. The degree of correlation as a function of the time delay between the detected photons measured in the rectilinear (a), diagonal (b), and circular bases (c). The degree of correlation is shown in (d) as a function of the rotation of the linear detection basis by a half-wave plate. The sinusoidal green dotted line shows simulates a perfect classically polarisation correlated source. The green dashed line shows the upper average limit for a classical source of photon pairs, and the grey dashed line shows the average expected for an unpolarised (e.g. white light) source.

photon. For classically polarisation correlated photon pairs, as observed previously in quantum dots with finite splitting[9-11], the degree of correlation varies sinusoidally with wave-plate angle[14], as shown schematically by the dotted line, contrary to the emission of this dot. In fact, the average linear correlation measured is 62.4 ± 2.4% for this dot, which is ten standard deviations above the 50% limit for classical pairs of photons, which proves that the quantum dot emits polarisation entangled photon pairs.

Finally, the entangled state $(|H_{XX}H_X\rangle+|V_{XX}V_X\rangle)/\sqrt{2}$ can also be written in the circular polarisation basis as $(|L_{XX}R_X\rangle+|R_{XX}L_X\rangle)/\sqrt{2}$. Therefore, polarisation anti-correlation should be observed in the co-circular polarised measurement basis. Such a correlation, measured between the R polarised XX and R and L polarised X photons, is shown in figure 2(c). A large degree of anti-correlation was measured, similar in magnitude, but opposite in sign to the degree of co-linearly polarised photons at -58%, consistent with the emission of entangled photon pairs.

To fully characterise the two photon state emitted by the dot, the two photon density matrix can be constructed from correlation measurements, using quantum state tomography[19]. Adapting the scheme for our measurements, it is possible to construct a density matrix using twelve measurements of the degree of correlation $C$. The measurements pairs required are the combinations of the V, H, L, and D biexciton polarisations, with the rectilinear, diagonal and circular polarised exciton detection bases. The resulting density matrix representing the emission from the quantum dot is shown in Figure 5, with real and imaginary components shown in (a) and (b) respectively.

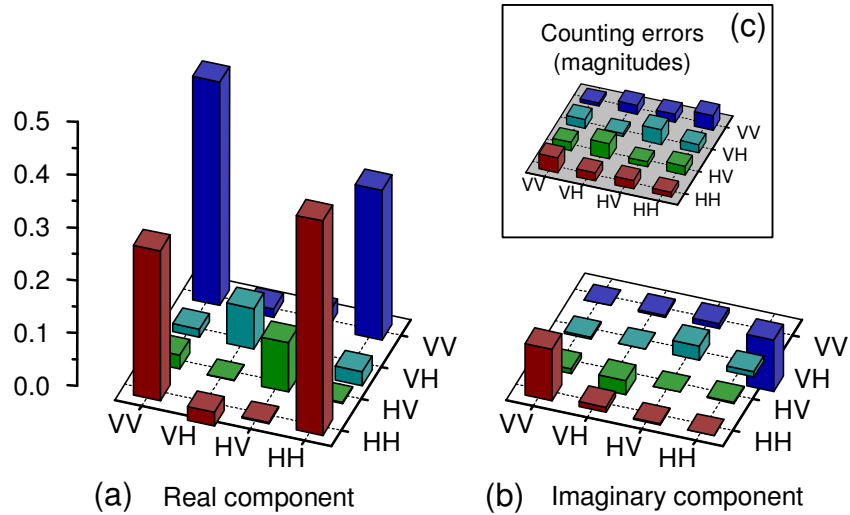

**Figure 5:** Real (a) and imaginary (b) components of the density matrix for a quantum dot with bright exciton splitting <1.5μeV. The inset (c) shows the magnitudes of the counting errors for the sixteen components of the density matrix. A scale for all elements, including the errors, is shown on the left hand side.

The strong outer diagonal elements in the real matrix demonstrate the high probability that the photon pairs have the same linear polarisation. The inner diagonal elements represent the probability of detecting oppositely linearly polarised photons, which is greatly suppressed over previous measurements. The residual average value of these elements is 0.085, of which we estimate 42% is due to background light. The remaining contribution is likely to be due to scattering of the exciton spin states. This too seems to be suppressed compared to previous measurements, though it is unclear if this is a direct result of reducing the density of wetting layer states resonant with the dots.

A direct consequence of the reduction of the background light is that the matrix more closely resembles the state of photon pairs generated by the dot itself. The outer off-diagonal elements in the real matrix are several times stronger than previously, and are clear indicators of entanglement. Small imaginary off-diagonal elements are additionally seen, which indicates that there may be a small phase difference between the |HH> and |VV> components of the entangled state. All other elements are close to zero, given the errors associated with the procedure, which are determined from the number of coincidences, and potted in figure 4(c).

The measured two photon density matrix projects onto the expected $(|H_{XX}H_X\rangle+|V_{XX}V_X\rangle)/\sqrt{2}$ state with fidelity $0.702 \pm 0.022$. This proves that the photon pairs we detect are entangled, since for pure or mixed classical un-polarised states, the fidelity cannot exceed 0.5. Numerous tests exist to prove that a quantum state is entangled, a selection of which we evaluate in Table 1. All of the tests are positive for entanglement, by many standard deviations. Perhaps the most interesting is the eigenvalue test, which determines the most probable state of the system. The result is approximately the maximally entangled state $(|H_{XX}H_X\rangle+e^{i(0.1\pi)}|V_{XX}V_X\rangle)/\sqrt{2}$, with eigenvalue $0.719\pm0.023$. For unpolarised classical light, the eigenvalue, or probability of emission into a specific polarisation state, cannot exceed 0.5.

| Test description | Test limit | Test result |
|---|---|---|
| $(|HH\rangle + |VV\rangle)/\sqrt{2}$ projection | >0.5 | 0.702 ± 0.022 |
| Largest eigenvalue | >0.5* | 0.719 ± 0.023 |
| Concurrence[19] | >0 | 0.440 ± 0.029 |
| Tangle[20] | >0 | 0.194 ± 0.026 |
| Average linear correlation | >0.5 | 0.624 ± 0.024 |
| Peres[21,**] | <0 | -0.219 ± 0.021 |

**Table 1:** Tests for entanglement performed on the density matrix of the quantum dot. The requirements to prove the state is entangled are quoted as the test limits. The test results are tabulated, along with errors determined by the finite number of counts in the experiment. All tests are positive for entanglement, with an average certainty of 9.5 standard deviations.

\* For an un-polarised source such as the one measured here.
\*\* For a state to be in-separable, the partial transpose must have at least one negative eigenvalue. We therefore measure the most negative eigenvalue of the partial transpose as the test quantity.

In conclusion, we directly demonstrate triggered emission of polarisation entangled photon pairs from a single quantum dot. By modifying the growth conditions we achieved a blue-shift of the wetting layer emission, which significantly reduces the background light intensity at the wavelength of quantum dots that emit close to 1.4 eV. The resulting improvements to the degree of polarisation correlation were equally significant, confirming background light to be a limiting factor to the observed degree of entanglement in degenerate quantum dots. The striking improvements in the quality of the light emission represents a break-through in the search for a useful and robust source of polarisation entangled photon pairs.

**Acknowledgements**

This work was partially funded by the EU projects SANDiE and QAP, and by the EPSRC.